\documentclass[reprint,amsmath,amssymb]{revtex4-1}

\usepackage{color,epsfig,rotating,graphicx}
\usepackage{amsmath,amssymb,amscd}

\usepackage[latin1]{inputenc}
\usepackage[english]{babel}

\usepackage{dsfont}
\usepackage{textcomp}
\usepackage{setspace}

\renewcommand{\Im}{{\rm Im}}

\newcommand{\ri}{{\rm i}}
\newcommand{\re}{{\rm e}}
\newcommand{\rd}{{\rm d}}

\newcommand{\rp}{{\rm p}}
\newcommand{\rs}{{\rm s}}

\newcommand{\kb}{k_{\rm B}}

\begin{document}
\author{S. Lang}
\affiliation{Institute of Optical and Electronic Materials, Hamburg University of Technology, 21073 Hamburg, Germany}
\email{slawa.lang@tuhh.de}

\author{M. Tschikin}
\affiliation{Institut f\"{u}r Physik, Carl von Ossietzky Universit\"{a}t, D-26111 Oldenburg, Germany.}

\author{S.-A. Biehs}
\affiliation{Institut f\"{u}r Physik, Carl von Ossietzky Universit\"{a}t, D-26111 Oldenburg, Germany.}

\author{A. Yu. Petrov}
\affiliation{Institute of Optical and Electronic Materials, Hamburg University of Technology, 21073 Hamburg, Germany}

\author{M. Eich}
\affiliation{Institute of Optical and Electronic Materials, Hamburg University of Technology, 21073 Hamburg, Germany}

\title{Large penetration depth of near-field heat flux in hyperbolic media}


\begin{abstract}
We compare super-Planckian thermal radiation between phonon-polaritonic media and hyperbolic metamaterials. In particular, we determine the penetration depth of thermal photons inside the absorbing medium for three different structures: two semi-infinite phonon-polaritonic media supporting surface modes, two multilayer hyperbolic metamaterials and two nanowire hyperbolic metamaterials. We show that for hyperbolic modes the penetration depth can be orders of magnitude larger than for surface modes suggesting that hyperbolic materials are much more preferable for near-field thermophotovoltaic applications than pure phonon-polaritonic materials.
\end{abstract}
\maketitle
\date{\today}
\makeatletter
\begin{center}
\@date
\end{center}
\makeatother


%
%


Theoretically it is well-known for a long time that heat radiation at the nanoscale can surpass the blackbody limit by orders of magnitude~\cite{PvH1971,ZhangReview2009,ZhangReview2013}. Recently, this super-Planckian property has been confirmed experimentally~\cite{HuEtAl2008,ShenEtAl2009,Kralik2012}. The underlying mechanisms resulting in elevated radiative heat fluxes are nowadays well understood. In particular, surface phonon polaritons can enhance the radiative heat flux by several orders of magnitude~\cite{JoulainEtAl2005} due to the very large number of contributing modes~\cite{BiehsEtAl2010}. However, it has been shown very recently that also so called hyperbolic modes~\cite{Nefedov2011,BiehsEtAl2012,GuoEtAl2012} can lead to an enormous increase in the radiative heat flux which can be even larger than that caused by surface modes~\cite{BiehsEtAl2012}. 

This effect of enhanced near-field thermal radiation has several possible applications~\cite{ZhangReview2009,ZhangReview2013} as touchless cooling~\cite{Ottens2011,GuhaEtAl2012} and near-field imaging~\cite{Yannick,KittelEtAl2008,HuthEtAl2011,WorbesEtAl2013} for instance. But probably one of the most discussed is near-field thermophotovoltaics (nTPV)~\cite{MatteoEtAl2001,NarayanaswamyChen2003,ParkEtAl2007,GuoEtAl2013,LarocheEtAl06} which exploits the near-field enhancement effect for increasing the output power and the efficiency of TPV devices. Structures using hyperbolic materials for nTPV were proposed recently~\cite{Nefedov2011,Simovski2013}.

Surface modes are supported by many different materials, like metals and polar materials. On the other hand hyperbolic modes only exist in very few natural materials like calcite and tetradymites~\cite{DrachevEtAl2013,ThompsonEtAl1998,EsslingerEtAl2014}. But it is possible to fabricate hyperbolic metamaterials (HMMs) which support hyperbolic modes. These HMMs can be constructed by combining metals or metal-like materials with dielectrics in subwavelength structures~\cite{PoddubnyEtAl2013,CaiEtAl2010}. The main advantage of the hyperbolic modes with respect to the surface modes is that they are propagating inside the hyperbolic medium~\cite{PoddubnyEtAl2013}, whereas the surface modes are bound to the surface of the material which is one of the bottlenecks of nTPV~\cite{ParkEtAl2007}.

The purpose of this letter is to analyze the penetration depth (PD) of thermal photons~\cite{BasuZhang2009,BasuZhang2011}, resulting from the heat exchange between two bodies close to each other. This quantity is very important for possible nTPV applications. It defines the effective thickness of the layer in which electron-hole pairs can be generated. But also for cooling applications larger PDs are preferable because photons transport heat much faster than phonons. Furthermore it is better to have a larger volume where the heat is absorbed to avoid local overheating.

%
%

Let us consider two half spaces at slightly different temperatures $T$ and $T+ \Delta T$ with $\Delta T \ll T$ separated by a vacuum gap with the width $l$. The half spaces consist of homogeneous, isotropic or uniaxial materials with optical axes parallel to the surface normal. Then the radiative heat flux through the gap is given by $\Phi = h_0 \Delta T$ where the heat transfer coefficient (HTC) is~\cite{PvH1971}
\begin{equation}
  h_0 =  \sum_{j = \rs,\rp} \int_0^\infty \!\!\! H_0^j(\omega) \frac{\rd \omega}{2 \pi}\\
\label{Eq:htc0}
\end{equation}
introducing the spectral heat transfer coefficient (sHTC)
\begin{equation}
  H^j_0 (\omega) = \frac{\rd \Theta(\omega,T)}{\rd T} \int_0^\infty \!\!\! \mathcal{T}^j(\omega,k_\rho) k_\rho \frac{\rd k_\rho}{2 \pi}.
\label{Eq:HTC0}
\end{equation}
Here $\Theta(\omega,T) = \hbar \omega / \bigl(\exp(\hbar \omega / \kb T) - 1\bigr)$ is the Bose-Einstein function, $\omega$ is the angular frequency, $k_\rho$ is the wave vector component of the radiation parallel to the interfaces, $\kb$ is the Boltzmann constant, $\hbar$ is the reduced Planck constant and $c$ is the speed of light in vacuum. The energy transmission coefficient $\mathcal{T}^j$ for s- and p-polarized waves is given by
\begin{equation}
   \mathcal{T}^j (\omega, k_\rho; l) = \begin{cases}
                \frac{(1 - |r^j_1|^2)(1 - |r^j_2|^2)}{|1 - r^j_1 r^j_2 \exp(2 {\rm i} \gamma_0 l)|^2}, & \!\! k_\rho < k_0 \nonumber \\
                \frac{ 4 \Im(r^j_1) \Im(r^j_2) \exp(-2 |\gamma_0| l) }{ |1 - r^j_1 r^j_2 \exp(-2 |\gamma_0| l)|^2 },  & \!\! k_\rho > k_0
  \end{cases}, 
\end{equation}
where $r^\rs$ and $r^\rp$ are the Fresnel coefficients for reflection of s- and p-polarized light at the half spaces 1 and 2 and $\gamma_0 = \sqrt{k_0^2 - k_\rho^2}$ is the z component of the wavevector inside the vacuum gap with $k_0 = \omega/c$.

%
%

\begin{figure}
  \epsfig{file = 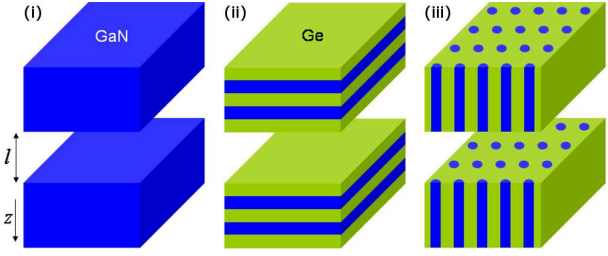, width = 0.45\textwidth}
  \caption{Systems to be analyzed. Two identical half spaces of (i) bulk GaN, of (ii) GaN/Ge layer
           HMMs and of (iii) GaN/Ge wire HMMs separated by a vacuum gap. The HMMs are modeled as
           effective media. The GaN filling factor of the layer HMM is 50 \%, the one of the wire HMM is 30 \%.
           The gap width is $l$. \label{Fig1:Structure}}
\end{figure}

In the following we consider the heat exchange by thermal radiation for the three different structures depicted in Fig.~\ref{Fig1:Structure}: (i) two gallium nitride (GaN) half spaces, (ii) two multilayer HMM (mHMM)~\cite{HoffmanEtAl2007,LangEtAl2013,KrishnamoorthyEtAl2012} half spaces composed of GaN/Ge bilayers and (iii) two nanowire HMM (wHMM)~\cite{YaoEtAl2008,NoginovEtAl2009} half spaces consisting of GaN nanowires immersed in a Ge host. The optical response of GaN is in the infrared dominated by the optical phonons so that the relative permittivity of GaN can be described by a Drude-Lorentz model~\cite{Adachi} $\epsilon_{\rm GaN} = \epsilon_\infty (\omega_{\rm LO}^2 - \omega^2 - \ri \omega_{\rm col} \omega)/(\omega_{\rm TO}^2 - \omega^2 - \ri \omega_{\rm col} \omega)$ with $\epsilon_\infty = 5.35$, $\omega_{\rm LO} = 141\cdot10^{12}\,{\rm rad/s}$, $\omega_{\rm TO} = 106\cdot10^{12}\,{\rm rad/s}$, and $\omega_{\rm col} = 1.52\cdot10^{12}\,{\rm rad/s}$. It is important to notice that inside the {\itshape reststrahlen band} of GaN ($\omega_{\rm TO} < \omega < \omega_{\rm LO}$) the permittivity is negative so that GaN can support surface phonon polaritons inside this frequency range. On the other hand, the permittivity of germanium (Ge) is in the infrared to very good approximation dispersionless having a value of $\epsilon_{\rm Ge} = 16$~\cite{OptSocAm}. Both materials are nonmagnetic.

%
%

In order to describe the optical response of the structures (ii) and (iii) we use effective medium theory (EMT) which gives reliable results if the unit-cell size of the underlying structure is much smaller than the wavelength. In the far-field regime the dominant wavelength is the thermal wavelength which is about $10\,\mu{\rm m}$ at $300\,{\rm K}$. In the near-field regime the contribution of the evanescent waves sets another constraint to the applicability of the effective description. As a rule of thumb the effective description gives reliable results for distances $l$ larger than the unit-cell size~\cite{BiehsEtAl2013,TschikinEtAl2013}. For smaller distances it tends to overestimate the hyperbolic heat flux contribution~\cite{LiuShen2013,BiehsEtAl2013}. According to the EMT the effective permittivity of the mHMM is given by~\cite{Yeh}
\begin{align}
  \epsilon_\rho = \epsilon_x = \epsilon_y &= f \epsilon_{\rm GaN} + (1 - f)\epsilon_{\rm Ge}, \\
  \epsilon_z &= \biggl( \frac{f}{\epsilon_{\rm GaN}} + \frac{1 - f}{\epsilon_{\rm Ge}} \biggr)^{-1},
\end{align}
and for the wHMM it is given by~\cite{WangbergEtAl2006}
\begin{align}
  \epsilon_\rho = \epsilon_x = \epsilon_y &= \epsilon_{\rm Ge} \frac{(1 + f) \epsilon_{\rm GaN} + (1 - f)\epsilon_{\rm Ge}}{(1 - f) \epsilon_{\rm GaN} + (1 + f)\epsilon_{\rm Ge}}, \\
  \epsilon_z &= f \epsilon_{\rm GaN} + (1 - f) \epsilon_{\rm Ge},
\end{align}
where $f$ is the volume filling fraction of GaN. We choose $f$ = 50 \% for the mHMM and $f$ = 30 \% for the wHMM. z is the direction perpendicular to the surfaces. Note that although we call the structures (ii) and (iii) HMMs they are hyperbolic only in certain frequency ranges. For the mHMM the range is $106-141\cdot10^{12}\,{\rm rad/s}$ and for the wHMM it is $106-121\cdot10^{12}\,{\rm rad/s}$ (with a small non-hyperbolic region from $111-112\cdot10^{12}\,{\rm rad/s}$). In this paper only effective media are considered. Real structures are to be studied in later publications.

%
%

Now, we want to define the PD $\delta$ of the thermal radiation into the colder medium. To this end, we need to determine the heat flux or the heat transfer coefficient inside the colder medium. It is already clear that due to the rotational symmetry of the problem the heat flux is along the z direction only. Furthermore, the components of the electric and magnetic fields parallel to the interfaces are continuous which implies that the z component of the Poynting vector is continuous at the interface as well. Finally, we know that the intensity of a plane wave with a given frequency $\omega$ and tangential wavevector $k_\rho$ impinging on a semi-infinite uniaxial medium (with the optical axis along the surface normal) is damped by a factor $\exp(- 2 \Im(\gamma^j) z)$ where $\gamma^j$ is the polarization dependent wavevector component along z direction, i.e. the direction of propagation of the heat flux. Hence, from Eq.~(\ref{Eq:HTC0}) giving the spectral heat transfer coefficient inside the vacuum gap and in particular at the interface of the absorbing medium, we can easily deduce the sHTC inside the colder medium. We obtain
\begin{equation}
       H^j (\omega; z) = \frac{\rd \Theta(\omega,T)}{\rd T} \int_0^\infty \!\!\! \mathcal{T}^j(\omega,k_\rho) \re^{-2 \Im(\gamma^j) z} k_\rho \frac{\rd k_\rho}{2 \pi}.
\end{equation}
assuming the vacuum / cold medium interface is located at z = 0. By replacing $H^j_0$ by $H^j$ in Eq.~(\ref{Eq:htc0}) we obtain the heat transfer coefficient $h(z)$ inside the colder medium. The z component of the wavevector is in our case given by
\begin{equation}
  \gamma^\rs = \sqrt{k_0^2 \epsilon_\rho- k_\rho^2 }
\end{equation}
for s-polarized waves ({\itshape ordinary waves}) and by
\begin{equation}
  \gamma^\rp = \sqrt{k_0^2 \epsilon_\rho - k_\rho^2 \frac{\epsilon_\rho}{\epsilon_z} }
	\label{Eq:gammaP}
\end{equation}
for p-polarized waves ({\itshape extra-ordinary waves}). Obviously, the anisotropy makes itself felt through the p-polarized waves. For $\epsilon_\rho \epsilon_z < 0$  the p-polarized waves have hyperboloidal isofrequency curves in k-space~\cite{PoddubnyEtAl2013}. The above introduced mHMM and wHMM support broad frequency bands for such hyperbolic modes. 

The PD $\delta$ is now defined such that it determines the distance at which the heat transfer coefficient $h$ has dropped to $1/\re$ of its value inside the vacuum gap, i.e.\ $h(\delta) = h(0)/\re = h_0/\re$. In a similar way we define the spectral PD $\delta(\omega)$ as the distance at which the spectral heat transfer coefficient $H = H^\rs + H^\rp$ drops to $1/\re$, i.e.\ $H(\omega,\delta(\omega)) = H(\omega,0)/\re = H_0(\omega)/\re  = (H^\rs_0(\omega)+H^\rp_0(\omega))/\re$.

%
%
\begin{figure*}[Hhbt]
  \epsfig{file = 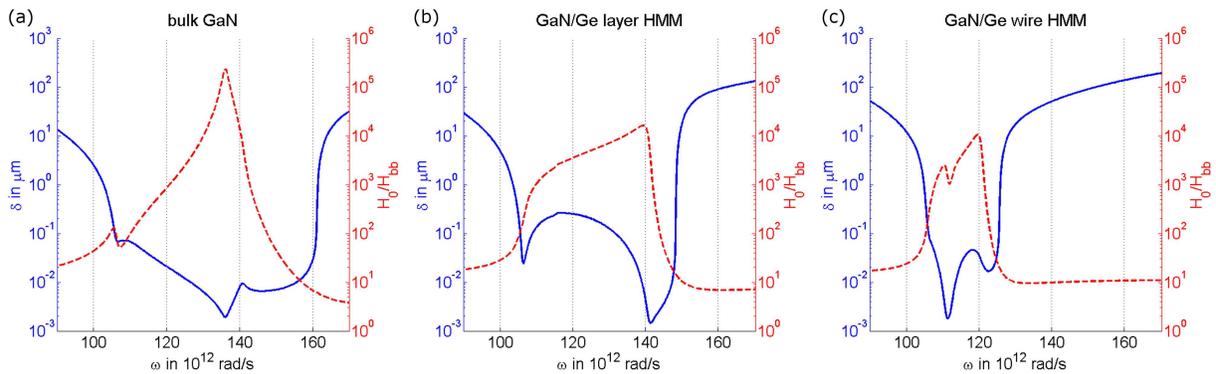, width = 0.9\textwidth}
  \caption{Spectral thermal PD (solid) and sHTC in the vacuum gap (dashed) for (a) bulk GaN, (b) GaN/Ge layer HMMs and (c) GaN/Ge wire HMMs. The sHTC is normalized to the sHTC between black bodies. The gap width is $l = 10\,{\rm nm}$.
           \label{Fig2:sHTC}}
\end{figure*}

Before discussing the total PD for the different systems (i)-(iii) we present in Fig.~\ref{Fig2:sHTC} the sHTC $H_0(\omega)$ and the spectral PD $\delta(\omega)$ for the different structures choosing $l = 10\,{\rm nm}$ and $T = 300\,{\rm K}$. It can be seen in Fig.~\ref{Fig2:sHTC}(a) that the heat flux between bulk GaN has a large, narrow peak at the surface mode frequencies as expected in this case~\cite{JoulainEtAl2005}. For the two HMMs in Fig.~\ref{Fig2:sHTC}(b) and (c) the heat flux is strong in the broad frequency bands where hyperbolic modes exist. These bands are due to the so called type I and type II responses~\cite{PoddubnyEtAl2013}. For the mHMM in Fig.~\ref{Fig2:sHTC}(b) both bands touch each other which is due to the choice of $f = 0.5$ resulting in a broad plateau for the sHTC and the spectral PD. For the wHMM in Fig.~\ref{Fig2:sHTC}(c) both bands are separated by a small, non-hyperbolic region (when choosing $f = 1/3$ instead of $f = 0.3$ both bands would touch as well). That is the reason for the small dip in the middle  of the plateau of the sHTC.

For all structures the spectral PD decreases when the sHTC increases. This is not surprising if Eq.~(\ref{Eq:gammaP}) is recalled. The larger $k_\rho$ the larger is the imaginary part of $\gamma^\rp \approx i k_\rho \sqrt{\epsilon_\rho / \epsilon_z}$ ($k_\rho \gg k_0$). And the large heat fluxes are caused by p-polarized, high $k_\rho$ modes. For the HMMs the imaginary part of $\gamma^\rp$ is due to the imaginary parts of $\epsilon_\rho$ and $\epsilon_z$ because in a lossless case this term $\sqrt{\epsilon_\rho / \epsilon_z}$ is just an imaginary number and the modes are not damped at all. But for GaN the term is 1 and damping is always observed independent of losses. Comparing the results in Fig.~\ref{Fig2:sHTC} for the different structures (i)-(iii) it becomes apparent that the PD in the HMMs can have values as small as the PD in GaN at the peak frequency of the sHTC. However, considering the whole hyperbolic region $\delta(\omega)$ can for the HMMs also have values which are one or even three orders of magnitude larger than the corresponding value for bulk GaN. By optimizing the HMMs to have smaller absorption this difference in PD can be made even larger.

%
%

\begin{figure}[Hhbt]
  \epsfig{file = 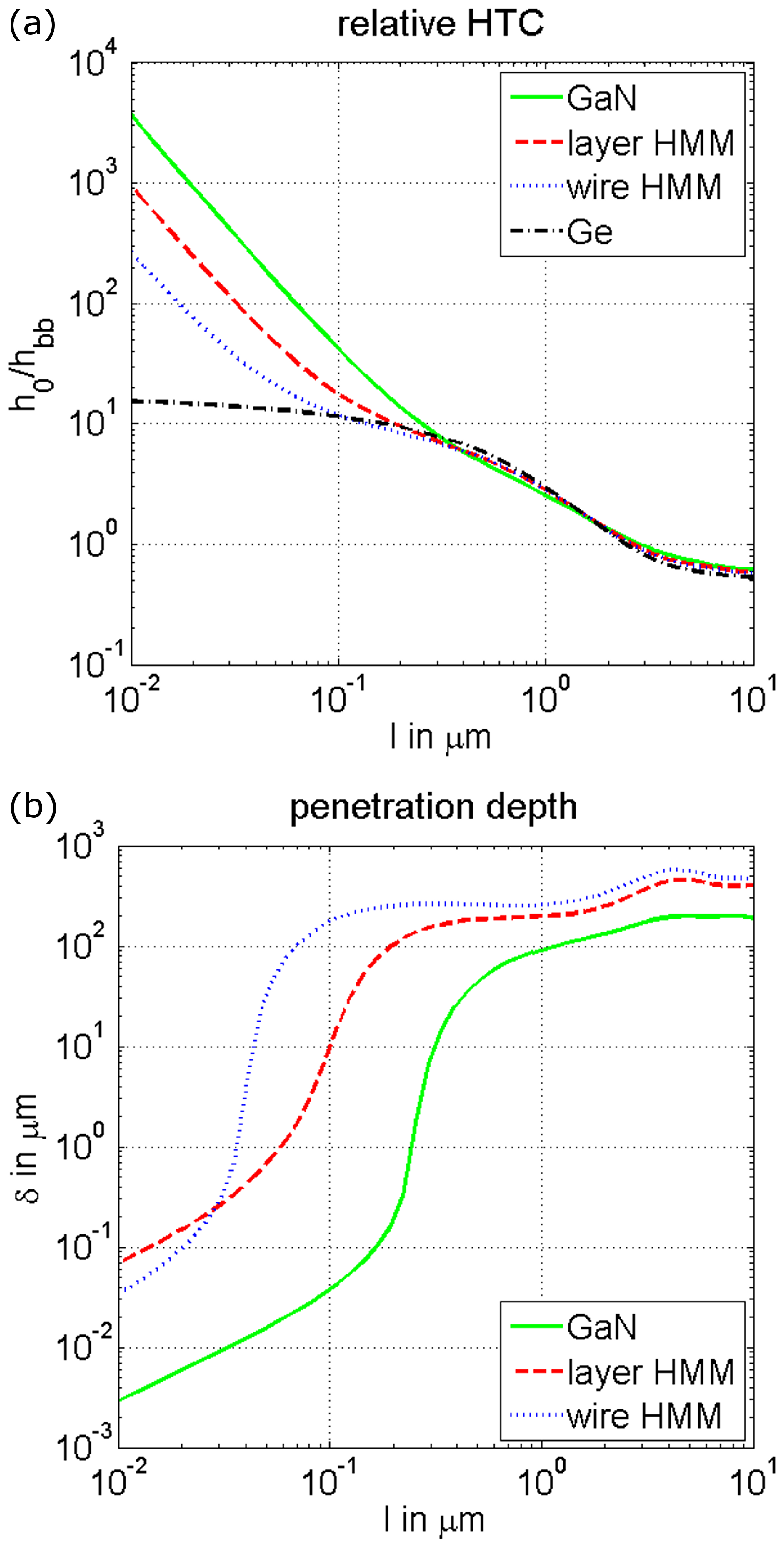, width = 0.45\textwidth}
  \caption{HTC in the vacuum gap (a) and thermal PD (b) at $T = 300\,{\rm K}$ for the systems: bulk GaN, GaN/Ge layer HMM, GaN/Ge wire HMM and bulk Ge. The HTC is normalized to the HTC between black bodies.
           \label{Fig3:PDandHTC}}
\end{figure}

In Fig.~\ref{Fig3:PDandHTC}(a) the total HTC for the three systems and additionally between two bulk Ge half spaces is shown. It is normalized to the HTC between two black bodies $h_{\rm BB}$ which is about $6.12\,{\rm W}/({\rm m}^2 {\rm K})$ independent of the gap width. The HMMs and GaN show the well known $1/l^2$ dependence in the near field. It should be emphasized that real HMMs have a cutoff in the k-space which depends on the unit-cell size of the structure resulting in a saturation of the heat flux for distances $l$ smaller than the unit-cell size~\cite{BiehsEtAl2013}. The relative heat flux of Ge approaches the value $\epsilon_{\rm Ge} = 16$ because the heat transfer is sustained by frustrated total internal reflection modes only which contribute by photon tunneling~\cite{Cravalho}. Since we have neglected losses for Ge its PD is infinite. Of course, real Ge has losses due to imperfections but the PD $\delta$ can still be very large. As can be observed in Fig.~\ref{Fig3:PDandHTC}(a) the wHMM has a relatively weak heat transfer coefficient compared to bulk GaN and the mHMM. However, we have checked that by replacing Ge with some dielectric having a lower refraction index one can make the hyperbolic band much broader and thus enhance the near-field heat flux significantly~\cite{BiehsEtAl2012}.

Finally, the total PD $\delta$ is depicted in Fig.~\ref{Fig3:PDandHTC}(b). In the far-field and the intermediate region it is more or less constant for all materials. The higher the filling factor of GaN the more lossy is the effective material and the shorter the penetration. To get an approximate boundary between far and near field we calculated the distance $l$ at which 50 \% of the total heat flux is due to hyperbolic modes. This distance is 115 nm for the mHMM and 44 nm for the wHMM --- so approximately given by the start of the $1/l^2$ dependence of the HTC. The spectral heat flux is generally much broader (e.g. for two black body half spaces) than the hyperbolic bands such that they only dominate in the near field on which our focus lies in this paper. In this strong near-field regime where the surface or hyperbolic modes dominate the heat flux for GaN and the HMMs, $\delta$ drops dramatically. This behavior can be understood by the fact that with smaller $l$ modes with $k_\rho$ on the order of $1/l$ dominate the thermal radiation so that $\Im(\gamma^j) \propto k_\rho \propto 1/l$ and hence the PD becomes approximately proportional to $l$. Most importantly in the intermediate regime the PD in the HMMs can be two to three orders of magnitude larger than in GaN and in the strong near-field regime it can be more than one order of magnitude larger than in GaN. Hence, our numerical results suggest that hyperbolic materials are preferable to phonon-polaritonic media when larger near-field PDs are needed as in the case of nTPV~\cite{ParkEtAl2007}.

It should be mentioned that the structures presented here are not directly applicable to nTPV systems. In this paper we report on the general property of hyperbolic modes supporting large penetration depths of near-field heat flux. For nTPV applications the materials should produce hyperbolic regions in the near infrared where photovoltaic cells are available. Concrete systems are left for further studies.

%
%

In conclusion, we have shown with concrete examples that the PD of thermal photons in the near-field regime can be much larger for materials or systems supporting hyperbolic modes than for materials supporting surface modes. The reason is the different nature of those modes. The penetration depth is potentially an important property for a variety of applications including touchless cooling and thermophotovoltaics.

The authors from Hamburg University of Technology gratefully acknowledge financial support from the German Research Foundation (DFG) via SFB 986 $"M^3"$, project C1. M. T. gratefully acknowledges support from the Stiftung der Metallindustrie im Nord-Westen. S.-A. B. and M. T. acknowledge financial support by the DAAD and Partenariat Hubert Curien Procope Program (project 55923991).

\end{document}